\documentclass[11pt]{article}
\usepackage[letterpaper,margin=1in]{geometry}

\usepackage{graphicx}
\DeclareGraphicsExtensions{.pdf,.png,.jpg,.jpeg}

\graphicspath{{figures/}{pictures/}{images/}{./}}

\usepackage{microtype}
\PassOptionsToPackage{warn}{textcomp}
\usepackage{textcomp}
\usepackage{mathptmx}
\usepackage{times}

\usepackage{cite}
\usepackage{booktabs}
\usepackage{listings}
\usepackage{xcolor}

\title{Domain-Grounded Tool Orchestration for LLM-Guided Scientific Analysis}

\author{
  Jeff Lee \quad Sebastien Jourdain \quad Cory Quammen \quad Patrick O'Leary \quad Berk Geveci\\[2pt]
  Kitware, Inc.\\
  \texttt{\{jeff.lee, sebastien.jourdain, cory.quammen, patrick.oleary, berk.geveci\}@kitware.com}
}
\date{}

\usepackage{hyperref}

\begin{document}

\maketitle

\begin{abstract}
Scientific analysis workflows encode deep domain knowledge through sequences of
tightly coupled operations where correctness depends on tool selection, execution
order, and parameterization. A CFD engineer investigating flow separation must
extract wall shear stress, identify zero-crossings in skin friction, and confirm
with boundary-layer profiles: a chain that requires both domain expertise and
proficiency with visualization tools. Current approaches to LLM-assisted scientific
visualization generate scripts that encode this knowledge implicitly, and often
incorrectly, producing code that executes but yields wrong results. We present an
architecture that separates intent interpretation (LLM) from execution
(deterministic domain tools) from explanation (LLM), connected by the Model Context
Protocol (MCP) and grounded by domain ontologies that constrain planning to valid
analysis chains. We instantiate the architecture in two domains on the same ParaView
server infrastructure: computational fluid dynamics post-processing and topological
data analysis via the Topology ToolKit (TTK). Adding the second domain required only
an ontology and tool wrappers around existing filters, with no change to the
architecture, protocol, or deployment. By construction the design removes whole
classes of failure that affect script generation (such as API hallucination and
missing pipeline stages) and narrows the strategic errors that remain. An ablation across both domains locates
the ontology's empirical effect: it does not change which tools the planner selects,
which is already reliable, but corrects how the model interprets results, raising
interpretation accuracy from 0.41 to 0.91, and only when the relevant fact is
retrieved in scoped rather than bulk form. ParaView's client-server model carries
analysis to production-scale datasets through a thin browser client.
\end{abstract}

\section{Introduction}

Scientific toolkits for data analysis and visualization are remarkably capable.
ParaView~\cite{Ahrens:2005:PEO} provides hundreds of filters for processing
simulation data.
The Topology ToolKit (TTK)~\cite{Tierny:2018:TTK} offers persistence diagrams,
Morse-Smale complexes, contour trees, and critical point extraction.
3D Slicer~\cite{Fedorov:2012:3DS} contains hundreds of modules for medical
image analysis.
The computational capability exists.
The gap is not in the tools but in their accessibility: given a scientific
question, which tools should be applied, in what order, with what parameters,
and what do the results mean?

Large language models (LLMs) offer a natural interface for bridging this gap.
A user can state an engineering question (``is there flow separation on the
upper surface?'') and an LLM can, in principle, translate that into the
appropriate sequence of analysis operations.
However, current approaches to LLM-assisted scientific visualization conflate
two fundamentally different tasks: deciding \textit{what} to analyze (an intent
interpretation problem where LLMs excel) and \textit{how} to execute the
analysis (a procedural domain knowledge problem where deterministic tools are
superior).

The dominant paradigm, LLM-generated script generation, asks the model to produce
executable code that directly invokes visualization APIs.
Recent work by Zhao et al.~\cite{Zhao:2026:RAG} demonstrates that even with
retrieval-augmented generation (RAG), this approach suffers from API
hallucination, missing pipeline stages, and wrong-strategy errors.
Their error taxonomy reveals that the hardest failures are not syntactic but
strategic: the LLM chooses the wrong analysis approach entirely, producing code
that executes but answers the wrong question.
These are domain knowledge failures, not code quality failures, and they cannot
be resolved by better code generation.

We propose an architecture that assigns each task to the component best suited
for it.
The LLM interprets user intent and explains results, tasks that require natural
language understanding and synthesis.
Deterministic domain tools execute the analysis, tasks that require validated,
reproducible procedures.
The Model Context Protocol (MCP)~\cite{MCP:2024} provides a standardized
interface between them.
A domain ontology constrains the LLM's planning to physically or topologically
valid analysis chains, preventing the wrong-strategy errors that persist in
script generation approaches.

We demonstrate that this pattern generalizes across scientific domains by
implementing it for two distinct applications on the same ParaView server
infrastructure: computational fluid dynamics (CFD) post-processing and
topological data analysis via TTK.
Both domains exhibit the same structural challenge: powerful tools that
require expert knowledge to configure, chain, and interpret. Both
benefit from the same architectural solution.
Adding the second domain required only an ontology file and MCP tool wrappers
around existing TTK filters; no changes to the architecture, protocol,
deployment, or user interface.

Our contributions are:
\begin{enumerate}
\item A \textbf{Plan--Execute--Interpret} architecture with domain ontology
and MCP tool layer that separates intent interpretation from execution from
explanation.
\item \textbf{CFD instantiation} with four use cases: validated analysis from
natural language, iterative multi-step investigation, cross-quantity synthesis,
and in-situ monitoring and control of a live Catalyst simulation.
\item \textbf{TTK instantiation} with three use cases: topological feature
identification, structural decomposition, and comparative topology across
timesteps.
\item Evidence that both domains \textbf{compose on the same server instance},
and that the marginal cost of adding a domain is an ontology and tool wrappers.
\item A \textbf{domain-independent in-situ capability}: the control and
monitoring loop demonstrated on a live CFD run (connect, pause/resume, step,
breakpoint, extract) applies to any Catalyst-instrumented HPC solver, with only
the monitored field names being domain-specific; a closed observe-reason-control
loop that one-shot script generation cannot provide.
\item An account of \textbf{why tool orchestration removes whole classes of
failure} (API hallucination and missing pipeline stages) by construction, and
narrows the strategic errors that remain via the ontology.
\item \textbf{Scalable deployment} via ParaView's client-server architecture
with a thin browser client.
\end{enumerate}

\section{Related Work}

\textbf{LLM-assisted visualization code generation.}
Several systems use LLMs to generate visualization scripts from natural language.
ChatVis~\cite{Mallick:2024:ChatVis,Peterka:2025:ChatVis} generates Python scripts
for ParaView, using chain-of-thought prompting and iterative error correction.
VizGenie~\cite{Biswas:2025:VizGenie} targets VTK directly, proposing self-refining
domain-aware workflows for scientific visualization.
Su et al.~\cite{Su:2025:NIST} describe an AI-powered ParaView system for NIST
research data.
Zhao et al.~\cite{Zhao:2026:RAG} investigate structure-aware RAG for vtk.js
pipeline generation, using module overlap rather than semantic similarity for
retrieval.
Specialized LLM systems have also been developed for specific visualization
tasks: Jeong et al.~\cite{Jeong:2024:TF} use natural language to generate
transfer functions for volume rendering, and Ai et al.~\cite{Ai:2025:VolVis}
present a multi-agent system for volumetric scene exploration and editing.
These approaches share a common architecture: the LLM generates executable code
that directly invokes visualization APIs.
The limitations are structural: generated scripts are one-shot, correction is
human-driven, and the most dangerous failure mode (code that executes but
performs the wrong analysis) is not addressed by better code generation.

\textbf{ParaView-MCP.}
Most closely related to our work, Liu et al.~\cite{Liu:2025:PVMCP} present
ParaView-MCP, an autonomous visualization agent that uses MCP for direct tool
invocation in ParaView.
Their system demonstrates that MCP is a viable protocol for LLM-driven
visualization.
Our work differs in three ways: (1)~we introduce domain ontologies that
constrain LLM planning to physically valid analysis chains, addressing the
wrong-strategy errors that tool invocation alone does not prevent;
(2)~our tools return structured quantitative data enabling a closed-loop
Plan-Execute-Interpret cycle where interpretation drives follow-up analysis;
and (3)~we demonstrate generalization across two scientific domains (CFD and
topological analysis) on the same server infrastructure.

\textbf{Agentic CFD systems.}
A growing body of work applies LLM-based agents to CFD workflows, particularly
around OpenFOAM.
OpenFOAMGPT~\cite{Pandey:2025:OFGPT} uses retrieval-augmented generation for
OpenFOAM case setup and analysis.
Its successor, OpenFOAMGPT 2.0~\cite{Feng:2025:OFGPT2}, introduces a
multi-agent framework with four specialized agents including a post-processing
agent for querying simulation outputs.
MetaOpenFOAM 2.0~\cite{Chen:2025:Meta} chains LLM reasoning for automated CFD
simulation and post-processing, with a separate post-processing agent (limited
to 2D plots and scalar queries).
OptMetaOpenFOAM~\cite{Chen:2025:Opt} extends this with sensitivity analysis and
parameter optimization.
Foam-Agent~\cite{Yue:2025:Foam} uses multiple FAISS indices for RAG across
OpenFOAM tutorials, scripts, and documentation.
Beyond CFD, InferA~\cite{Tam:2025:InferA} offers a multi-agent approach for
analyzing ensembles of cosmological simulations, including a visualization
component.
These systems demonstrate convergence toward agentic tool orchestration for
scientific analysis.
However, they remain solver-specific (primarily OpenFOAM), lack domain
ontologies for constraining analysis planning, and do not return structured
results for LLM-based interpretation.

\textbf{Natural language interfaces for scientific tools.}
Natural language interfaces for visualization have been explored extensively in
the chart-oriented domain~\cite{Luo:2022:NL2VIS,Li:2024:MatPlotAgent}, where the
mapping from intent to visual encoding is relatively direct.
Scientific visualization presents a harder problem: the mapping from an
engineering question to a multi-step analysis procedure requires domain
expertise that varies across disciplines, solvers, and data formats.

\textbf{Tool use and agentic architectures.}
The Model Context Protocol (MCP)~\cite{MCP:2024} standardizes how LLMs discover
and invoke external tools, providing typed parameters, structured return values,
and a stateless protocol over stateful backends.
Function calling and tool use in LLMs have been explored through
ReAct~\cite{Yao:2023:ReAct} and chain-of-thought
prompting~\cite{Wei:2022:CoT}.
Our work differs in that the tools return structured quantitative data, not
confirmation messages or rendered images, enabling the LLM to perform
quantitative reasoning in the interpretation phase.

\textbf{Domain ontologies in scientific computing.}
Ontologies for scientific simulation have been developed across
CFD, structural mechanics, and materials science.
Our contribution is an ontology specifically designed to ground LLM planning
in domain knowledge, mapping observable phenomena to the physical quantities
that reveal them and the analysis tools that extract those quantities.

\section{Architecture}
\label{sec:arch}

This section presents the domain-independent architectural pattern.
\autoref{sec:cfd} and \autoref{sec:ttk} instantiate it for CFD and
topological data analysis.

\subsection{System Overview}

\begin{figure}[htbp]
 \centering
 \includegraphics[width=\columnwidth]{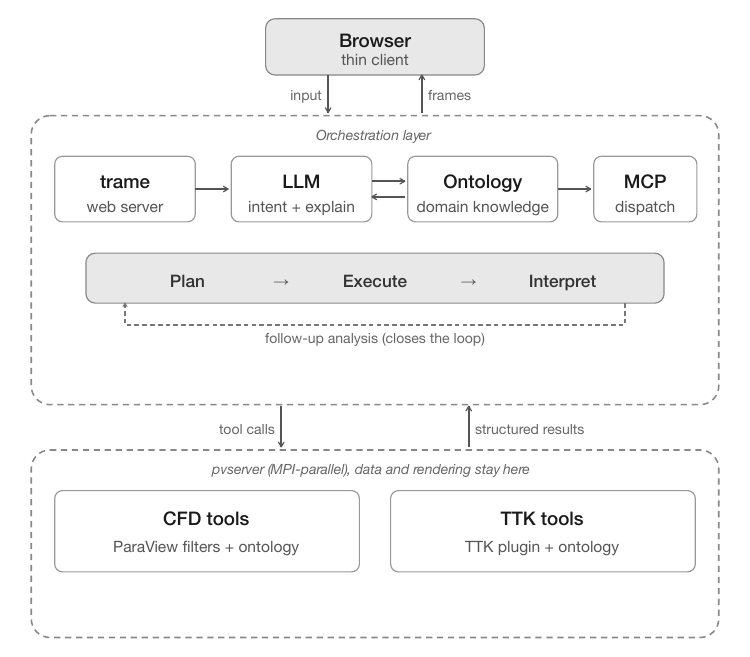}
 \caption{System architecture. The browser is a thin client receiving only
 rendered frames. The LLM plans analysis operations constrained by the domain
 ontology, dispatches them via MCP to deterministic tools running on
 \texttt{pvserver}, and interprets the structured results. Both CFD and TTK
 tools execute on the same server instance. The Plan--Execute--Interpret loop
 closes when the Interpret phase proposes follow-up analysis.}
 \label{fig:architecture}
\end{figure}

The system comprises four layers (\autoref{fig:architecture}):
(1)~a \textbf{browser client} that receives compressed rendered frames and
sends user input. No data or geometry reaches the browser; rendering is
performed entirely server-side via ParaView's \texttt{VtkRemoteView};
(2)~a \textbf{trame web server}~\cite{Jourdain:2023:trame} that mediates
between the browser and the ParaView backend, hosting the LLM integration
and MCP tool dispatch;
(3)~an \textbf{LLM with domain ontology} that interprets user intent, plans
analysis operations constrained by the ontology, and interprets structured
results returned by tools;
(4)~a \textbf{ParaView server} (\texttt{pvserver}) that holds all data,
executes the visualization pipeline, and renders frames, optionally running
MPI-parallel on a compute node with both CFD filters and TTK filters
available as ParaView plugins.

The same codebase runs in two modes: in-process via \texttt{pvpython} for
development, or client-server via \texttt{pvserver} for production.
A dual-mode data access helper transparently selects between in-process access
and client-server fetch based on the active connection type.

\subsection{The Plan--Execute--Interpret Loop}

Each user interaction initiates a cycle through three phases.

\textbf{Plan (Intent Resolution).}
The user states a question in domain language.
The LLM consults the domain ontology, which maps observable phenomena to the
physical quantities that indicate them and the tools that extract those
quantities.
The ontology \textit{constrains} the plan: rather than choosing freely from
all possible ParaView operations, the LLM selects from the subset that the
ontology declares valid for the identified phenomenon.
The output is an ordered sequence of named tool invocations with parameters.

\textbf{Execute.}
MCP dispatches the planned tool calls to the ParaView pipeline.
Tools are deterministic domain workflows, not generated scripts, that build
or modify the server-side pipeline and return structured results: field
statistics, extracted profiles, integrated quantities, topological invariants.
Critically, tools return \textit{data}, not \textit{renders}.
The pipeline persists across invocations, enabling multi-step analysis within
a single session.

\textbf{Interpret.}
The LLM receives the structured results along with the relevant ontology entry.
It generates a grounded explanation with quantitative specifics (``separation
detected at $x/c = 0.72$ based on $C_f$ sign reversal'') rather than generic
commentary.
The ontology's \texttt{follow\_ups} field provides diagnostic reasoning chains:
if separation is detected, the ontology recommends confirming with boundary
layer profiles and checking for upstream transition.
When the interpretation identifies a finding that warrants further investigation,
it proposes a follow-up analysis, closing the loop back to the Plan phase.

\subsection{Domain Ontology}

The ontology follows a domain-independent schema with five fields per entry:
\textit{phenomenon} (an observable feature or condition),
\textit{indicators} (quantities that reveal it),
\textit{tools} (operations that extract those quantities),
\textit{follow-ups} (what to investigate next),
and \textit{significance} (why it matters in the application context).
The schema is instantiated per domain. The CFD ontology follows it closely
across roughly 30 entries covering external aerodynamics. The TTK ontology
adapts it: it names the main topological feature types (persistence diagram,
critical points, the Morse-Smale complex, and simplification), each with an
interpretation and a note on why it matters, and adds a short set of analysis
recommendations that fix the tool ordering for common goals such as feature
identification. Each ontology is authored by a domain expert in JSON and loaded
at startup.

\autoref{fig:ontology} shows representative entries from the CFD and TTK
ontologies.
The ontology is deliberately small, capturing the high-leverage knowledge that
distinguishes expert analysis from naive tool application, rather than
attempting encyclopedic domain coverage.

\begin{figure}[htbp]
\centering
\begin{lstlisting}[basicstyle=\scriptsize\ttfamily,frame=single,
  caption={Representative ontology entries.},label={fig:ontology},
  captionpos=b,abovecaptionskip=5pt]
// CFD ontology entry
{ "phenomenon": "flow_separation",
  "indicators": ["Cf_zero_crossing",
    "reversed_BL_velocity"],
  "tools": ["show_forces", 
    "show_boundary_layer"],
  "follow_ups": ["boundary_layer_profiles",
    "upstream_transition_check"],
  "significance": "Increases pressure drag,
    reduces lift, may indicate stall onset"
}

// TTK ontology: a feature entry
{ "persistence_diagram": {
  "description": "birth-death pairs from
    TTKPersistenceDiagram",
  "persistence_gap": "a gap in the sorted
    persistence separates signal/noise",
  "design_relevance": "count of high-
    persistence pairs = significant
    features, independent of mesh"
}}

// TTK ontology: an analysis recommendation
{ "feature_identification": {
  "goal": "find significant features,
    separate them from noise",
  "chain": ["compute_persistence",
    "compute_simplification",
    "compute_critical_points"]
}}
\end{lstlisting}
\end{figure}

\subsection{MCP Tool Layer}

Tools are organized in two tiers.
\textbf{Domain workflow tools} encode expert knowledge about multi-step
analysis procedures.
In CFD, \texttt{show\_boundary\_layer} extracts a velocity profile normal to
the surface at a specified station and computes displacement thickness,
momentum thickness, and shape factor.
In TTK, \texttt{show\_persistence\_diagram} runs the persistence diagram filter
and returns the birth-death pairs with their types and persistence values.
These tools encapsulate the procedural knowledge that script generation handles
unreliably.
\textbf{Refinement tools} provide low-level control: changing color maps,
adjusting thresholds, modifying representation types.
These are domain-independent and shared across both instantiations.

Three design principles govern tool implementation.
\textit{Structured returns:} every tool returns quantitative data, not just
confirmation, data the interpretation phase can reason about.
\textit{Dynamic field resolution:} tools accept semantic field names
(``velocity,'' ``pressure'') and resolve them to solver-specific array names
at runtime using pattern matching and alias tables.
\textit{Safe pipeline management:} consumer-aware deletion prevents breaking
downstream filter dependencies, and infrastructure filters are automatically
reused when a matching instance already exists.

\subsection{Client-Server Deployment}

ParaView's client-server architecture provides scalability without
architectural changes.
The \texttt{pvserver} process runs on a compute node, holds all data in server
memory, executes the full visualization pipeline, and renders frames using
server-side GPUs.
The trame web server connects via \texttt{simple.Connect()} and forwards all
pipeline operations through ParaView's client-server protocol.
The browser receives only compressed JPEG frames. It never sees geometry or
field data.
Client memory remains constant regardless of dataset size; a 100-million-cell
mesh resides entirely on the server.
For analysis operations requiring local computation, a \texttt{fetch\_data}
helper transfers only extracted results (surfaces, profiles, integrated
quantities).

\section{Domain 1: Computational Fluid Dynamics}
\label{sec:cfd}

\subsection{Implementation}

The CFD instantiation wraps ParaView's visualization filters as MCP tools with
aerodynamics domain knowledge.
The ontology covers external aerodynamics phenomena including flow separation,
stagnation, recirculation, vortex breakdown, shock waves, and wake structure.
Each phenomenon maps to indicator quantities ($C_f$, $C_p$, wall shear stress,
velocity gradients, Q-criterion, total pressure loss) and the tools that
extract them.
The tool set includes
\texttt{show\_vortices} (Q-criterion or $\lambda_2$ isosurfaces),
\texttt{show\_streamlines} (seeded by wake extent detection),
\texttt{show\_boundary\_layer} (line probes normal to surface with integral
thickness computation),
\texttt{show\_forces} (pressure and viscous force integration with lift/drag
decomposition),
\texttt{show\_slice} (planar cuts with automatic field selection),
and \texttt{show\_flow\_topology} (vortex core line extraction).
Evaluation uses FUN3D and OpenFOAM datasets for external aerodynamics
configurations including airfoils, wings, and bluff bodies.

\begin{figure}[htbp]
 \centering
 \includegraphics[width=\columnwidth]{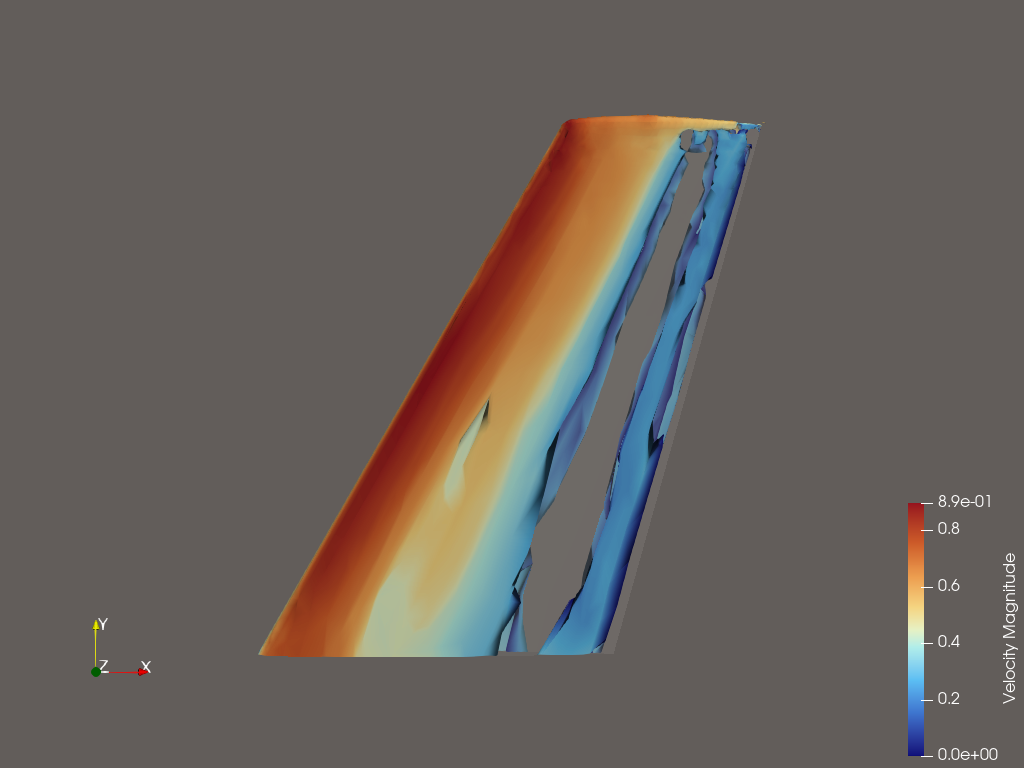}
 \caption{Q-criterion isosurfaces generated by the \texttt{show\_vortices}
 tool, colored by velocity magnitude. The tool automatically computes
 Q-criterion, selects a threshold based on field statistics, and configures
 the display. The LLM selected this tool based on the ontology's mapping
 from ``vortex structures'' to Q-criterion extraction.}
 \label{fig:vortices}
\end{figure}

\subsection{Use Case: Natural Language to Validated Analysis}

The three use cases below are representative walkthroughs that illustrate the
interaction pattern; the quantitative values shown (separation locations, shape
factors, persistence values) are illustrative of the reasoning loop rather than
measured benchmark results.

A user asks: ``Is there flow separation on the upper surface?''

In the \textbf{Plan} phase, the LLM consults the CFD ontology.
The entry for ``separation'' identifies $C_f$ zero-crossing and reversed
velocity in the boundary layer as indicators, and maps these to
\texttt{show\_forces} (for skin friction extraction) and
\texttt{show\_boundary\_layer} (for velocity profile inspection).
The planner produces a two-step analysis chain: (1)~extract skin friction
coefficient on the upper surface, (2)~locate zero-crossings.

In the \textbf{Execute} phase, the tools run deterministically.
The surface is extracted, the skin friction field is resolved dynamically
(mapping ``skin friction'' to \texttt{Cf} in FUN3D, \texttt{wallShearStress}
in OpenFOAM), and zero-crossings are identified programmatically.
The tool returns structured data: a list of $x/c$ locations where $C_f$
changes sign, along with the $C_f$ distribution as an array.

In the \textbf{Interpret} phase, the LLM receives this structured data and the
ontology's significance entry for separation.
It produces: ``Separation detected at $x/c = 0.72$ on the upper surface based
on $C_f$ sign reversal.
Reattachment occurs at $x/c = 0.85$.
Recommend boundary layer profiles upstream and downstream to characterize the
separation bubble structure.''
The follow-up recommendation comes from the ontology's \texttt{follow\_ups}
field, not from the LLM's training alone.

\subsection{Use Case: Iterative Diagnostic Reasoning}

This use case demonstrates the closed-loop nature of the architecture.

\textbf{Step 1.} The user asks: ``Show me the pressure distribution.''
The planner maps this to surface $C_p$ extraction and color-mapped display.
After execution, the interpreter examines the $C_p$ distribution and identifies
a plateau on the upper surface between $x/c = 0.55$ and $x/c = 0.70$.
The ontology links $C_p$ plateaus to likely laminar separation bubbles
and recommends boundary layer profiles upstream and downstream.
The system reports: ``$C_p$ plateau detected on the upper surface between
$x/c = 0.55$--$0.70$, consistent with a laminar separation bubble.
Recommend boundary layer profiles at $x/c = 0.50, 0.60, 0.70, 0.80$.''

\textbf{Step 2.} The user approves the follow-up.
The planner invokes \texttt{show\_boundary\_layer} at four stations.
Each call returns a velocity profile array, displacement thickness $\delta^*$,
momentum thickness $\theta$, and shape factor $H$.
The interpreter synthesizes: ``Reverse flow detected at $x/c = 0.60$ and $0.70$,
confirming laminar separation.
Shape factor $H$ peaks at 4.2 at $x/c = 0.65$.
Attached profile recovered by $x/c = 0.75$ ($H = 1.8$), confirming turbulent
reattachment.
Separation bubble extends from $x/c \approx 0.57$ to $x/c \approx 0.73$.''

This multi-step diagnostic chain mirrors expert CFD practice.
Script generation cannot support it: scripts are one-shot, with no mechanism
for structured results to drive subsequent analysis.

\begin{figure}[htbp]
 \centering
 \includegraphics[width=\columnwidth]{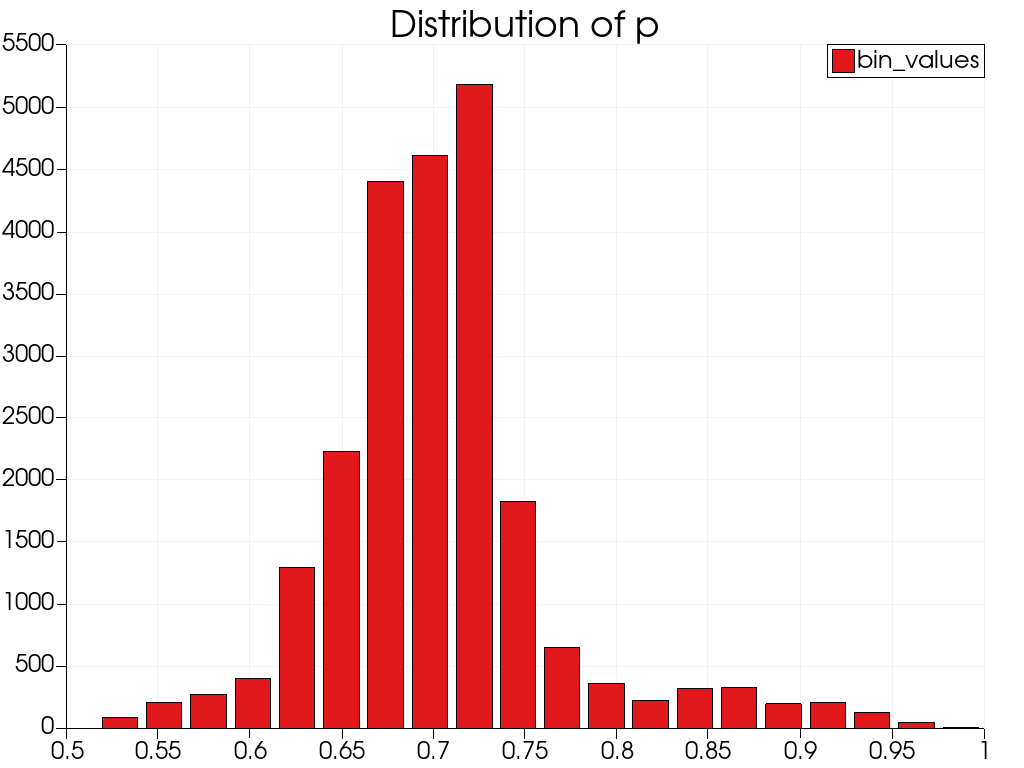}
 \caption{Pressure field distribution generated by \texttt{show\_histogram}.
 Tools return structured data (bin counts, statistics) that the Interpret
 phase uses for quantitative reasoning, not just rendered images.}
 \label{fig:histogram}
\end{figure}

\subsection{Use Case: Cross-Quantity Synthesis}

A user asks: ``What is driving the drag increase between $4^\circ$ and
$8^\circ$ angle of attack?''

The ontology's entry for drag identifies three components: pressure drag (from
$C_p$ integration), viscous drag (from wall shear integration), and wake
momentum deficit.
The planner decomposes the question:
(1)~compute pressure and viscous drag components for both angles via
\texttt{show\_forces},
(2)~extract separation location for both cases,
(3)~compare wake momentum deficit profiles.
All extractions execute within a persistent pipeline session; both datasets
remain loaded, and tools operate on each in turn.

The interpreter synthesizes: ``Separation moved forward from $x/c = 0.78$ at
$4^\circ$ to $x/c = 0.62$ at $8^\circ$.
Pressure drag increased 40\% while viscous drag remained roughly constant.
Wake momentum deficit increased 55\%.
Conclusion: separation-induced pressure drag dominates the drag increase,
consistent with incipient stall.''
No single tool call answers this question; the value is in orchestrated
multi-step reasoning where the ontology tells the planner how drag decomposes.

\begin{figure}[htbp]
 \centering
 \includegraphics[width=0.48\textwidth]{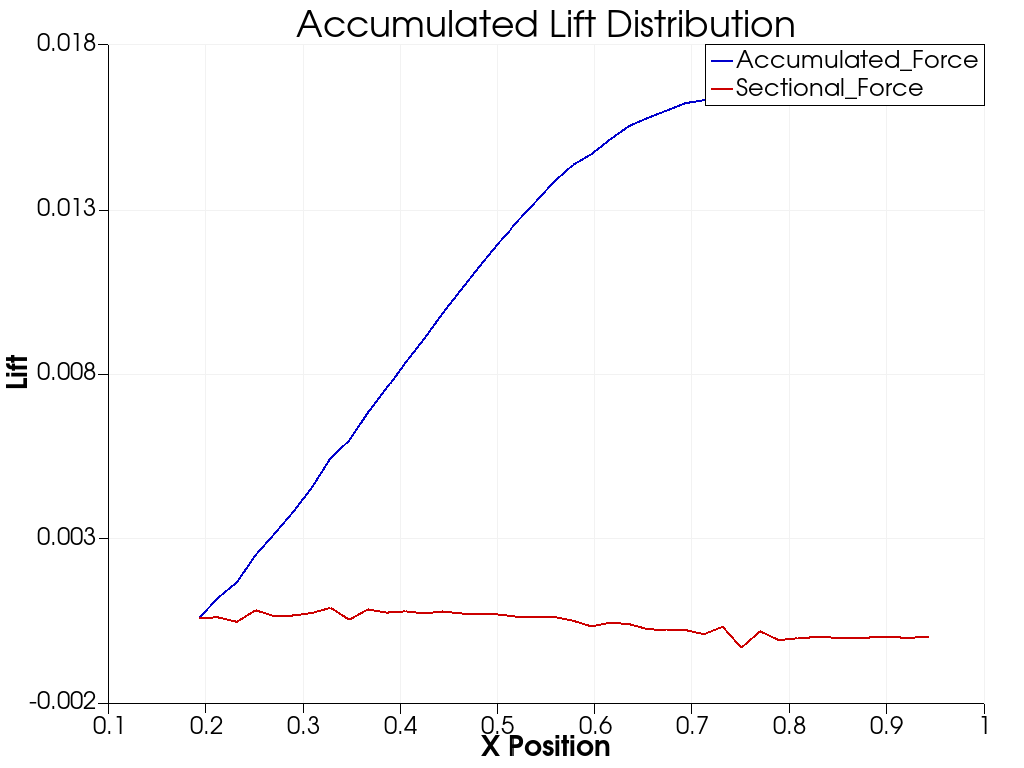}
 \hfill
 \includegraphics[width=0.48\textwidth]{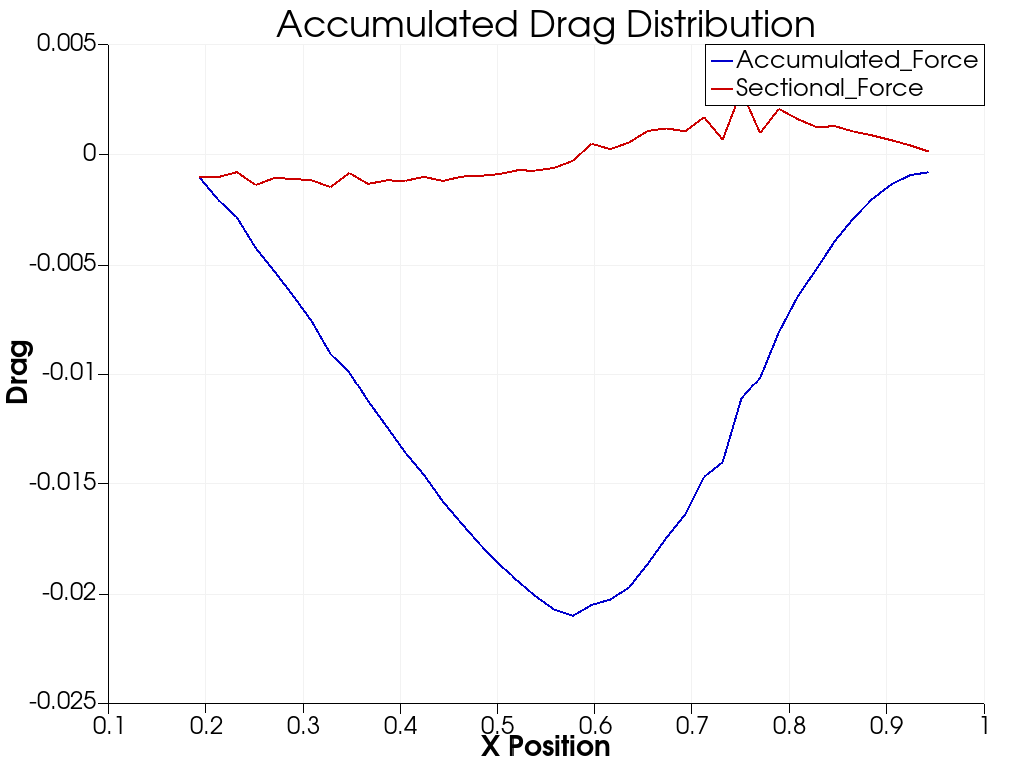}
 \caption{Force distributions generated by \texttt{show\_forces} for the
 cross-quantity synthesis use case. Left: lift distribution. Right: drag
 distribution. The tool integrates pressure and viscous contributions over
 the surface and returns structured decomposition data that the Interpret
 phase synthesizes into a physical explanation of the drag increase.}
 \label{fig:forces}
\end{figure}

\subsection{Use Case: In-Situ Monitoring and Control of a Live Simulation}
\label{sec:insitu}

The use cases above operate on saved results, but the same tool layer operates on
a running solver. When started with live mode enabled, the system connects to a
Catalyst-instrumented simulation through ParaView's Catalyst Live
protocol~\cite{Ayachit:2015:Catalyst}: \texttt{connect\_live} attaches to the
in-situ session while preserving the local rendering session, so that every
analysis tool acts on the live pipeline exactly as it does on static data. This
use case was exercised against a live Catalyst-instrumented FUN3D run.

\textbf{Plan.} The user works in execution-control language rather than
file-analysis language: ``connect to the simulation and pause it,'' ``step
forward 50 iterations,'' ``pause when the continuity residual stops dropping,''
``show me the surface pressure right now.'' The planner maps these to the live
control and monitoring tools.

\textbf{Execute.} The tool layer exposes deterministic control over the running
solver. \texttt{pause\_simulation} and \texttt{resume\_simulation} toggle the
run state; \texttt{step\_simulation} advances a fixed number of iterations and
pauses; \texttt{set\_breakpoint} arranges a pause when a target iteration is
reached. \texttt{extract\_monitors} subscribes to the solver's residual channel
and accumulates an iteration-indexed convergence history on the client
(continuity, the three momentum components, and energy), which
\texttt{show\_monitor\_chart} renders as a live log-scale plot.
\texttt{extract\_live\_data} copies a named source (a surface or a volume) out of
simulation memory into the local pipeline, where every CFD tool from the
preceding use cases (\texttt{show\_forces}, \texttt{show\_boundary\_layer},
\texttt{show\_slice}, and the rest) applies without modification. Control,
monitoring, and analysis all run against the same paused state.

\textbf{Interpret.} The residual history returned by the monitoring tools is
structured data, so the interpretation phase reasons over convergence the same
way it reasons over a $C_f$ distribution: it can report that the continuity
residual has flattened near $10^{-4}$ while the momentum residuals continue to
fall, recommend pausing to inspect the current solution, or set a breakpoint and
let the run proceed to it. Because extracted live fields enter the same pipeline
as static data, a diagnosis formed on a paused solution (a developing separation
on the upper surface, say) can be confirmed with the same boundary-layer and
force tools used on saved results.

This is a closed observe-reason-control loop over the simulation's execution
state, and it is qualitatively beyond what one-shot script generation can do: a
generated script is dispatched once and cannot interactively run, pause, monitor,
and re-inspect a live solver under the guidance of its own intermediate findings.
The loop here governs \emph{when} the simulation advances and \emph{what} is
extracted from it; steering of the solver's own inputs is the natural next step,
discussed in Section~\ref{sec:future}.

\section{Domain 2: Topological Data Analysis (TTK)}
\label{sec:ttk}

\subsection{Implementation}

TTK is a ParaView plugin; it runs on the same \texttt{pvserver} instance as
the CFD filters with no additional infrastructure.
Adding topological analysis required authoring a TTK-specific ontology and
implementing MCP tool wrappers around existing TTK filters.
The ontology names the topological features a user asks about (signal versus
noise features, the dominant extrema, the separatrix skeleton, and the
persistence gap that distinguishes signal from noise) and records the analysis
ordering those features imply, for example that the field must be simplified
before critical points are extracted or before the Morse-Smale complex is
segmented.
Indicator quantities include persistence values and critical point counts by
type (minima, 1-saddles, 2-saddles, maxima).

The tool set is four MCP tools, each wrapping one TTK filter:
\texttt{show\_persistence\_diagram} (TTKPersistenceDiagram, the birth-death
pairs that expose the signal-versus-noise gap),
\texttt{show\_critical\_points} (TTKScalarFieldCriticalPoints, extrema and
saddles colored by critical type),
\texttt{show\_morse\_smale} (TTKMorseSmaleComplex, the separatrix skeleton with
its critical points),
and \texttt{show\_simplified} (TTKTopologicalSimplificationByPersistence, which
cancels pairs below a persistence threshold to denoise the field).
Each tool takes a field name and operates on whatever dataset is loaded in the
shared pipeline, so the same four tools apply to a synthetic test field or to a
CFD scalar without modification.

\subsection{Use Case: Feature Identification and Denoising}

\begin{figure}[htbp]
 \centering
 \includegraphics[width=\columnwidth]{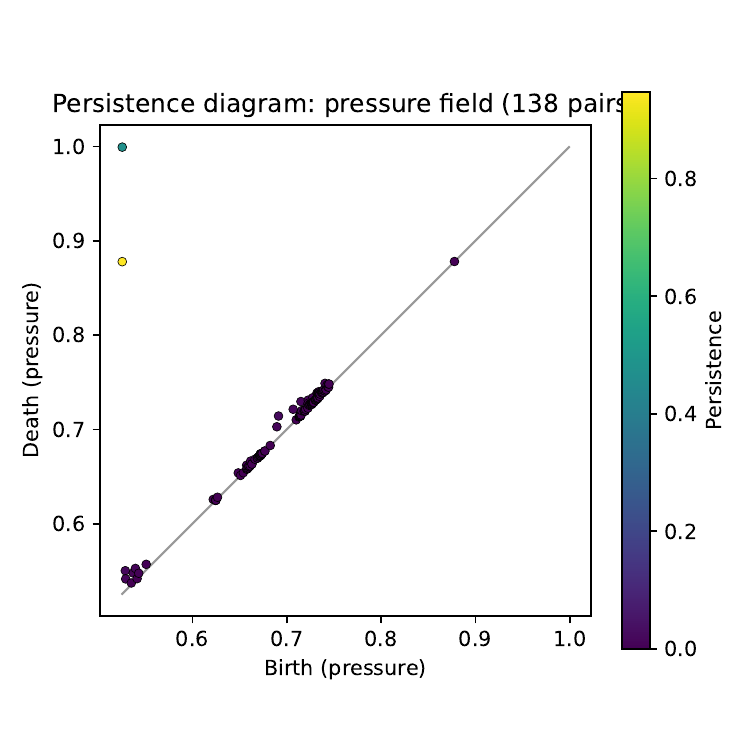}
 \caption{Persistence diagram of the FUN3D wing pressure field at one timestep,
 produced by \texttt{show\_persistence\_diagram}. Each point is a birth-death
 pair; distance above the diagonal is persistence. The cluster near the diagonal
 is discretization noise, and the few high-persistence pairs are the dominant
 pressure features. 138 pairs total.}
 \label{fig:persistence}
\end{figure}

\begin{figure}[htbp]
 \centering
 \includegraphics[width=\columnwidth]{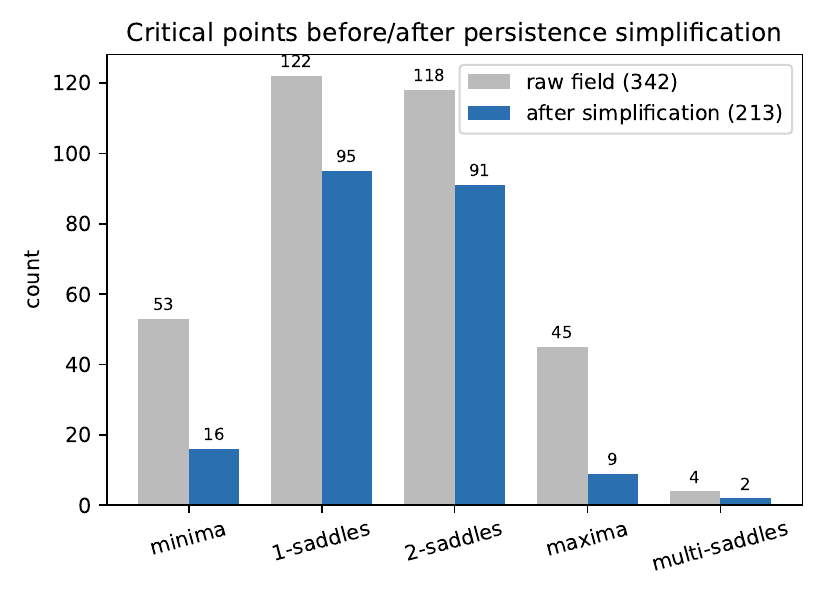}
 \caption{Critical-point counts before and after persistence simplification on
 the same field. Simplification at the gap threshold removes the low-persistence
 saddles and minima that correspond to mesh-scale noise while the high-persistence
 extrema survive, cutting the total from 342 to 213. The multi-saddles are
 degenerate saddles from the piecewise-linear discretization, explained in the
 text.}
 \label{fig:denoise}
\end{figure}

Unlike the illustrative CFD walkthroughs, the values below are measured: they
come from running the implemented tools on the pressure field of a FUN3D
simulation (a wing in transonic flow) at a single timestep, on the reassembled
serial mesh of roughly 16{,}000 points described in
Section~\ref{sec:crossdomain}.

A user asks: ``What are the dominant features in this scalar field?''

The ontology maps ``dominant features'' to a prerequisite chain: compute the
persistence diagram, identify the persistence gap that separates signal from
noise, simplify below the gap, then extract critical points of the simplified
field.
This ordering matters: extracting critical points on the unsimplified field
returns hundreds of points dominated by discretization noise.
Without the ontology, an LLM might skip simplification entirely or choose an
arbitrary threshold.

\textbf{Plan:} (1)~\texttt{show\_persistence\_diagram} on the pressure field,
(2)~identify the persistence gap from the returned distribution,
(3)~\texttt{show\_simplified} at a threshold inside the gap,
(4)~\texttt{show\_critical\_points} on the simplified field.

\textbf{Execute:} The persistence diagram returns 138 birth-death pairs
(\autoref{fig:persistence}). Retaining roughly the 25 most persistent pairs
corresponds to an absolute persistence threshold near 0.0038.

\textbf{Interpret:} On the raw field, critical-point extraction returns 342
points: 53 minima, 122 1-saddles, 118 2-saddles, 45 maxima, and 4 multi-saddles.
After simplifying at the gap threshold this falls to 213 (16 minima, 95
1-saddles, 91 2-saddles, 9 maxima, 2 multi-saddles), a 38\% reduction
concentrated in the low-persistence saddles and minima produced by mesh-scale
noise; the small set of high-persistence maxima (suction peaks) and minima
persists (\autoref{fig:denoise}). The retained extrema and the saddles linking
them are the dominant pressure structures over the wing.

The multi-saddles deserve a word, since they are a property of the
discretization rather than of the flow. On a smooth (Morse) field every critical
point is one of four types by Morse index: a minimum, a 1-saddle, a 2-saddle, or
a maximum. The pressure field here is piecewise linear on a tetrahedral mesh, and
TTK classifies each vertex from the connectivity of its lower and upper links
rather than from a Hessian. A vertex whose link is more complex than a simple
saddle, the discrete analogue of a monkey saddle with three ascending and three
descending sectors instead of two, is reported as a multi-saddle: one degenerate
point that stands in for several simple saddles. Two implications follow. First,
for the Euler-characteristic bookkeeping (minima minus 1-saddles plus 2-saddles
minus maxima) to balance, a multi-saddle must be counted with its multiplicity,
not as a single point; TTK can instead unfold each one into simple saddles by
symbolic perturbation when a strict Morse complex is required. Second,
multi-saddles arise where vertex values are nearly tied or the mesh is too coarse
to separate adjacent saddles, so they are low-persistence by nature: here they
are 4 of 342 points, and simplification removes half of them. We report them so
the counts are exact, but they carry no aerodynamic meaning of their own.

\subsection{Use Case: Structural Decomposition}

\begin{figure}[htbp]
 \centering
 \includegraphics[width=\columnwidth]{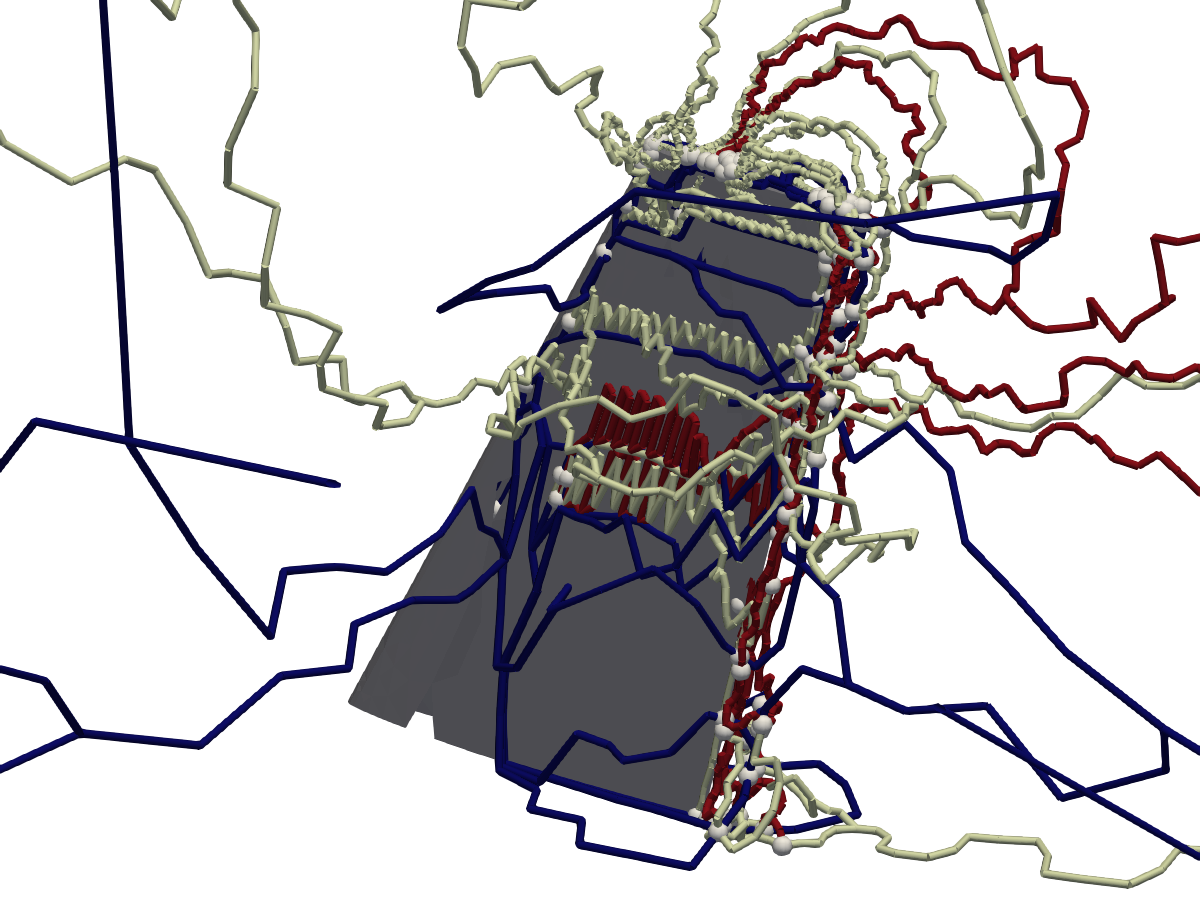}
 \caption{Morse-Smale separatrix skeleton and critical points of the wing
 pressure field, produced by \texttt{show\_morse\_smale} after simplification.
 The skeleton connects the surviving extrema through the saddle network.}
 \label{fig:morsesmale}
\end{figure}

A user asks: ``Show me the basin structure of this flow.''

The ontology maps ``basin structure'' to Morse-Smale decomposition, with a
prerequisite: the scalar field must be simplified before segmentation to avoid
producing thousands of spurious cells from numerical noise.
This prerequisite chain (simplify then segment) parallels the CFD
pattern (extract surface then compute wall shear): both encode ordering
constraints that domain experts know implicitly but that LLMs handle unreliably.

\textbf{Plan:} (1)~\texttt{show\_persistence\_diagram} on the scalar field,
(2)~\texttt{show\_simplified} at the identified noise threshold,
(3)~\texttt{show\_morse\_smale} on the simplified field.

\textbf{Execute:} \texttt{show\_morse\_smale} returns the critical points, the
separatrix skeleton connecting them, and the segmentation into ascending and
descending manifolds (\autoref{fig:morsesmale}).

\textbf{Interpret:} The interpretation below is illustrative of how the planner
narrates a segmentation rather than a measured breakdown of basin volumes:
``Morse-Smale decomposition reveals 4 ascending basins and
3 descending basins.
The largest ascending basin corresponds to the freestream.
Two smaller basins mark the primary and secondary recirculation zones.
Separatrices align with the observed shear layers, confirming that the
topological segmentation captures the physically meaningful flow structure.''

\subsection{Use Case: Comparative Topology}

\begin{figure}[htbp]
 \centering
 \includegraphics[width=\columnwidth]{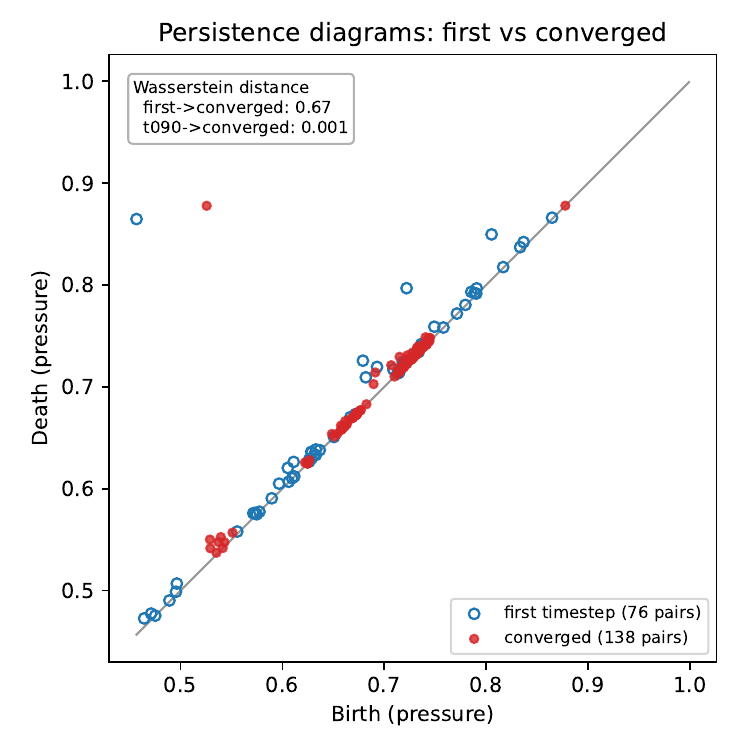}
 \caption{Persistence diagrams of the pressure field at the first timestep (open
 circles, 76 pairs) and at the converged state (filled, 138 pairs). As the
 solution develops, mid-persistence pairs fill in. The Wasserstein distance
 between the two diagrams is 0.67, while between a near-final timestep and the
 converged state it is 0.001, so the topology has stabilized.}
 \label{fig:compare}
\end{figure}

A user asks: ``How does the feature structure change between the first and last
timestep?''

The ontology maps ``change in feature structure'' to a comparison of persistence
diagrams: compute the diagram at each timestep, measure the distance between them,
and see which pairs appear or disappear.

\textbf{Plan:} (1)~\texttt{show\_persistence\_diagram} at the first and last
timesteps, (2)~measure the distance between the two diagrams with TTK's bottleneck
filter, (3)~overlay them for visual comparison (\autoref{fig:compare}).

\textbf{Execute (measured):} At the first timestep the pressure field has 76
birth-death pairs and 239 critical points, with pressure in [0.457, 0.980]. At the
last timestep it has 138 pairs and 340 critical points, with pressure in
[0.526, 0.999]. The Wasserstein distance between the two diagrams is 0.67, with the
matching cost split across minimum-saddle (0.45), saddle-saddle (0.33), and
saddle-maximum (0.37) pairs. Between timestep 90 and the last timestep the distance
is 0.001.

\textbf{Interpret:} This run is a steady-state computation, so the comparison
measures convergence rather than unsteady dynamics. The pressure topology develops
during the transient: as the suction peak and stagnation region organize, the
field gains persistent features and the diagram nearly doubles its pair count. By
timestep 90 the diagram has stopped changing (distance 0.001 to the final state),
so the persistence diagram acts as a topology-based convergence indicator. On a
time-accurate unsteady run the same two tools would instead track the birth and
death of features over physical time, for example across a shedding cycle; the
machinery is identical and only the interpretation of the distance changes.

The distance here is computed by applying TTK's bottleneck and Wasserstein filter to
the two diagrams. Wrapping it as a fifth topology tool would be as incremental as the other four,
the marginal-cost argument of this section applied once more.

\subsection{Why Quantitative Topology}

Persistence-based analysis is mature in scientific visualization, with a track
record in combustion, cosmology, materials, and turbulence studies, but it is
largely inaccessible without specialized expertise: a practicing CFD engineer reaches for
iso-surfaces, Q-criterion, line probes, and visual inspection because those are in
the GUI, while the topological tools are not in normal reach. The contribution
here is not the mathematics but the grounding, an ontology plus a tool layer that
lets these methods be invoked routinely on solver output rather than assembled as
one-off research artifacts. Each capability above answers a concrete need.

\textbf{Reproducible feature inventories.} Feature identification with persistence
answers ``how many distinct suction peaks are real, and which are mesh noise,''
replacing a hand-picked iso-value with a data-driven persistence gap and turning a
visual judgment into a reproducible count. The value is highest where no human is
in the loop: a design-of-experiments sweep of hundreds of runs, where each field
needs an automatic and defensible feature count rather than an analyst inspecting
every case.

\textbf{Structure extraction.} The Morse-Smale decomposition extracts the skeleton
of a field (separation and attachment lines, the saddle network) that an engineer
would otherwise trace by hand off surface streamlines. As an automated step it
yields a consistent structural description that can be compared across a parameter
sweep or across solvers, on the structure itself rather than on rendered pixels.

\textbf{Convergence and change detection.} The diagram distance introduced above
serves two purposes. On a steady run it is a topology-based convergence monitor: it
reports when the feature structure of the field has stopped changing, which is
closer to what an engineer cares about than an equation residual and can disagree
with it (a residual can plateau while a feature still drifts, or the topology can
settle before the last digits of the residual do). On a time-accurate run the same
distance tracks the birth and death of features over physical time, for example a
shedding cycle.

\textbf{Scoring AI surrogates.} The strongest use is as an objective for the
machine-learning side of the same workflow. A surrogate trained on pointwise error
will readily blur a shock or merge two vortices, because the pointwise penalty for
doing so is tiny and standard error metrics do not catch it. A persistence-diagram
distance gives a feature-aware signal instead: as a validation metric (``the
surrogate reproduces the field's critical-point structure to within a Wasserstein
distance of X'') or as a training regularizer that penalizes getting the feature
count and salience wrong. This closes the loop in our setting, where the same
persistence and distance machinery built for human-facing analysis becomes the
quantity that scores a learned model. For training use, the Wasserstein-2 distance
between diagrams is preferred over the bottleneck (L-infinity) distance because it
is smoother and has a usable gradient; differentiable persistence and diagram
vectorizations such as persistence images are an active research area that make
this practical by moving the per-step cost into a fixed feature space. We do not
claim a trained surrogate here. The point is narrower: the quantity such training
needs is exactly the one the analysis loop already produces.

\section{Cross-Domain Analysis}
\label{sec:crossdomain}

\subsection{What Transfers}

\autoref{tab:transfer} summarizes the division between domain-independent and
domain-specific components.
The architecture, protocol, deployment infrastructure, and user interface are
entirely shared.
Adding TTK required no modifications to any of these layers.
The domain-specific work consisted of authoring the ontology (a compact JSON
file) and implementing MCP tool wrappers that configure and invoke existing
TTK ParaView filters.

\begin{table}[tb]
  \caption{Domain-independent vs.\ domain-specific components.}
  \label{tab:transfer}
  \centering
  \scriptsize
  \begin{tabular}{ll}
  \toprule
  \textbf{Reused as-is} & \textbf{Domain-specific} \\
  \midrule
  Plan--Execute--Interpret loop & Ontology content \\
  MCP protocol and tool discovery & Tool implementations \\
  Ontology schema & Interpretation grounding \\
  Client-server deployment & Evaluation criteria \\
  trame UI and LLM integration & \\
  Safe pipeline management & \\
  Dual-mode data access & \\
  In-situ control and extraction & Monitor channel fields \\
  \bottomrule
  \end{tabular}
\end{table}

\subsection{Domain Independence of In-Situ Operation}

The in-situ capability of Section~\ref{sec:insitu} obeys the same split. Nothing
in the control loop is specific to fluid dynamics: connecting to a live Catalyst
session, pausing and resuming the solver, stepping by a fixed number of
iterations, setting an iteration breakpoint, and extracting a named source into
the pipeline are operations on a simulation's execution state, not on its physics.
Any HPC solver instrumented with Catalyst exposes the same control surface, so the
same control and extraction tools drive a structural, combustion,
molecular-dynamics, or climate solver without modification. The only
domain-specific element is the set of field names carried on the monitor channel:
the CFD instantiation accumulates continuity, momentum, and energy residuals,
whereas another domain would publish its own convergence or diagnostic quantities
(species residuals, displacement norms, energy drift). That mapping is authored
per domain exactly like an ontology entry, at the same marginal cost as the rest
of a domain instantiation. In-situ monitoring and control is therefore not a CFD
feature of the system but a domain-independent capability of the architecture,
parameterized by a short list of monitored quantities.

\subsection{Combined Workflows: TTK on CFD Data}

\begin{figure}[htbp]
 \centering
 \includegraphics[width=\columnwidth]{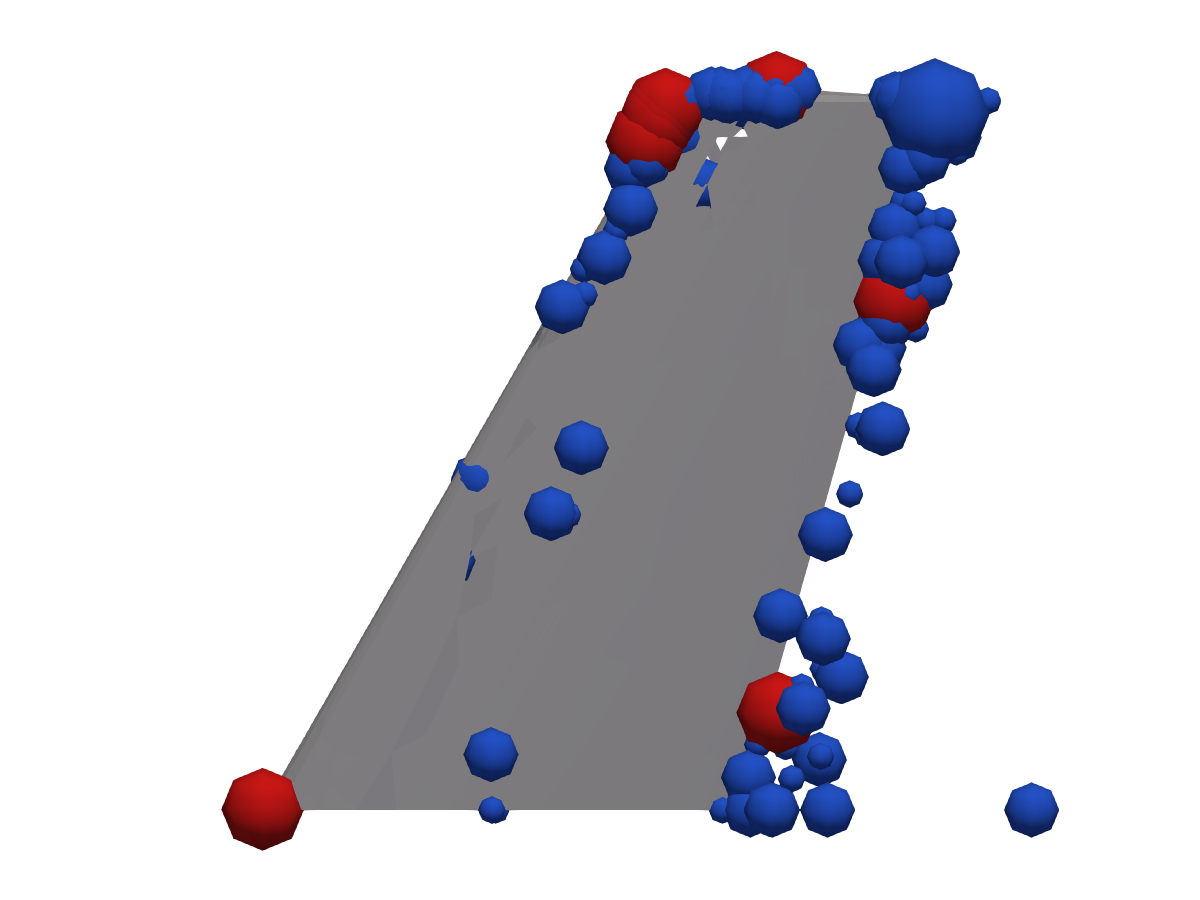}
 \caption{Critical points of the wing pressure field, colored by critical type,
 shown over the wing geometry. These are produced by the topology tool
 \texttt{show\_critical\_points} and can seed a CFD tool, for example streamlines,
 within the same session.}
 \label{fig:critical}
\end{figure}

Because both tool groups run in one MCP server (38 tools in the configuration
used here) on the same \texttt{pvserver} instance, they compose within a single
session: a user investigating a CFD dataset can invoke topology tools on a CFD
scalar field and feed the result back into a CFD tool, without switching systems
or transferring data.

We exercised this with a local model (a 32B code model served through Ollama,
not a frontier API) and a single cross-domain prompt asking for the pressure
field's critical points and streamlines seeded from them. From the 38 tools the
planner produced one plan that composed across the two domains:
\texttt{show\_critical\_points} on the pressure field (topology), then
\texttt{show\_streamlines} seeded from those critical points and colored by
velocity magnitude (CFD), then a screenshot (\autoref{fig:critical}). The
composition was correct on the first plan; the two domains did not need to be
wired together by hand, because the planner sees both tool groups and both
ontologies in one surface.

Execution surfaced two integration issues worth reporting, both consistent with
the paper's thesis rather than against it. First, the planner guessed the field
name ``Pressure'' when the array is named \texttt{p}, so the first
critical-points call returned nothing until the name was grounded against the
dataset: exactly the array-naming brittleness that Section~\ref{sec:noscripts}
attributes to script generation, and the reason field resolution belongs in the
tool layer rather than in a guessed string. Second, the seeding call exposed a
ParaView API change (the stream tracer's seed source moved to a dedicated
\texttt{StreamTracerWithCustomSource} filter in version 6.1), which the tool
wrapper now handles. Neither was a failure of the cross-domain reasoning.

Applying topology to CFD data also requires correct mesh bookkeeping. The FUN3D
output is written per rank, and each rank carries a halo of ghost cells that
duplicate cells owned by neighboring ranks. Concatenating the ranks and merging
coincident points is not enough: the duplicate cells remain and produce spurious
critical points along the partition interfaces. The reassembly used for the
measured results above therefore appends the ranks, removes ghost information,
and only then merges points, which for the timestep shown reduces the mesh from
117{,}503 cells (with ghosts) to 90{,}892 cells on about 16{,}000 points. The
architecturally cleaner answer for production is to stay distributed in situ and
let ParaView and TTK handle ghosts natively rather than gathering to a serial
mesh at all.

This cross-domain composition emerges from the architecture: the planner selects
tools from whichever ontology fits the query, and all tools operate on the same
server-side pipeline.

\subsection{Topology-Guided Deep Dive}

\begin{figure}[htbp]
 \centering
 \includegraphics[width=\columnwidth]{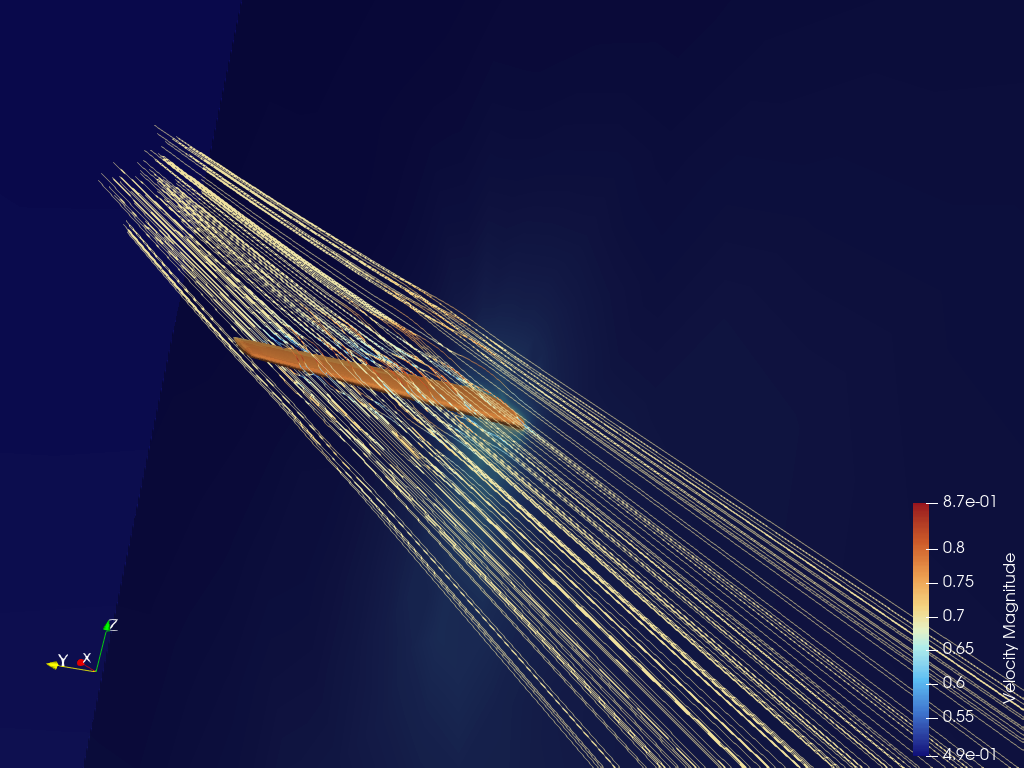}
 \caption{Wake-seeded streamlines (colored by velocity magnitude) over the wing,
 produced by \texttt{show\_streamlines} in the same session. The seeds are placed
 automatically in the wake region returned by the wake-detection tool; the
 streamlines fan downstream of the trailing edge, and the low-speed wake deficit
 trails the airfoil.}
 \label{fig:deepdive_scene}
\end{figure}

\begin{figure}[htbp]
 \centering
 \includegraphics[width=0.82\columnwidth]{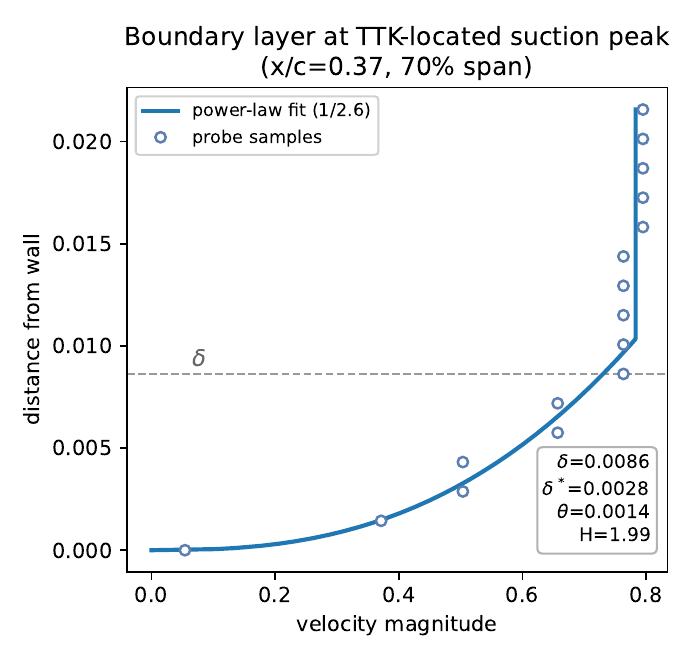}
 \caption{Boundary-layer velocity profile (probe samples and a power-law fit) at the
 chord station TTK flagged as the suction peak, x/c=0.37, 70\% span. Only three to
 four mesh cells span the layer, so the profile is coarsely resolved: the fit and the
 integral quantities are illustrative of the tool output, not a wall-resolved
 measurement (see text).}
 \label{fig:deepdive}
\end{figure}

\begin{figure}[htbp]
 \centering
 \includegraphics[width=\columnwidth]{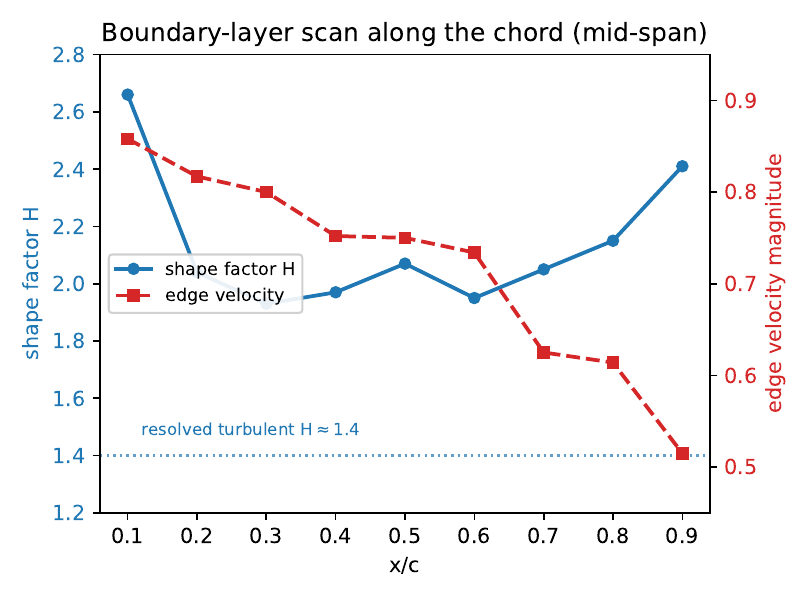}
 \caption{Boundary-layer scan along the chord at mid-span. The edge velocity (right
 axis) falls monotonically as the flow decelerates toward the trailing edge, a
 resolution-robust signal. The shape factor (left axis) stays near 2 everywhere and
 rises aft, well above the H$\approx$1.4 of a resolved turbulent layer: the boundary
 layer is turbulent but under-resolved on this mesh, the aft rise reflecting pressure
 recovery.}
 \label{fig:deepdive_trend}
\end{figure}

The combined workflow above used topology to seed a CFD visualization. The stronger
pattern is to let topology decide \emph{where} an expensive, localized CFD analysis
is applied: the persistence diagram and critical points identify a region of
interest, and the CFD tools then characterize it. This mirrors how an analyst
actually works, scan the field for something interesting and then probe it, with
the scan made quantitative and automatic.

We ran this end to end on the converged FUN3D field. TTK extracted the pressure
critical points and returned the lowest-pressure minimum, the suction peak, at
x/c=0.37 and 70\% span (the deepest of 53 minima, at p=0.53). That location was
handed to the CFD boundary-layer tool, which cut a section there
(\autoref{fig:deepdive}) and returned a boundary-layer thickness of 0.0086 chord and
a shape factor of about 2. In the same session the wake-detection tool located the
wake centroid just aft of the trailing edge (x=0.96 against a trailing edge at
x=0.95), and streamlines were seeded automatically in that wake region
(\autoref{fig:deepdive_scene}).

The shape factor is the most diagnostic quantity in this chain. A value near 2 is neither a
resolved turbulent profile (H$\approx$1.4) nor a clean laminar one (H$\approx$2.6),
which is ambiguous on its own. Cross-checking the Spalart-Allmaras eddy-viscosity
field resolves the ambiguity: the model is fully active across the domain (mean working variable
about 22, turbulent over 91\% of cells), so the layer is turbulent and the near-wall
mesh is too coarse to resolve it. Only three to four cells span the layer, and
the first off-wall sample sits at roughly a third of the boundary-layer thickness,
far coarser than a resolved turbulent profile needs. So H$\approx$2 here is a
resolution artifact, and the honest conclusion is under-resolved turbulent flow. That
is itself a useful result: the tools reach it quickly and quantitatively, and it
indicates which quantities are trustworthy.

A chord-wise scan (\autoref{fig:deepdive_trend}) supports this interpretation. The edge
velocity falls steadily from 0.86 near the leading edge to 0.52 near the trailing
edge; because it does not depend on the near-wall resolution, that deceleration is
trustworthy and marks the pressure-recovery region. The shape factor stays near 2 at
every station, never approaching the turbulent 1.4, and climbs in the aft third as
the boundary layer thickens under the adverse pressure gradient. The trends are the
resolution-robust signal here; the absolute shape factor is not.

None of these CFD tools know anything about topology, and the topology tools know
nothing about boundary layers. They compose because they share the pipeline and the
planner sees both tool groups: topology supplies a location, the CFD tools consume
it.

\subsection{Marginal Cost of Adding a Domain}

The TTK domain instantiation required a compact domain ontology (a 60-line JSON
file naming four feature types and the analysis ordering they imply) and about
400 lines of Python wrapping four TTK ParaView filters as four MCP tools, split
into a low-level analysis module and a thin workflow module that renders the
results. Wiring the new tools into the existing MCP server took only a few lines:
the topology tools are registered alongside the CFD tools and the topology
ontology is merged into the shared ontology.
No changes to the architecture, protocol, deployment, user interface, or LLM
integration were required.
The effort to add a new domain is proportional to the number of domain-specific
tools, not to the complexity of the architecture.
This suggests that the pattern can extend to additional scientific domains at
similar marginal cost, provided that the analysis tools already exist as
ParaView plugins or can be wrapped as MCP endpoints.

\section{Why Not LLM-Generated Scripts}
\label{sec:noscripts}

The alternative architecture is simpler: the user describes what they want, the
LLM generates a Python script, ParaView executes it.
This is the approach taken by most LLM-for-visualization
systems~\cite{Mallick:2024:ChatVis,Biswas:2025:VizGenie,Zhao:2026:RAG}.
We argue it is the wrong abstraction for scientific analysis.

\textbf{Brittleness across data sources.}
Generated scripts embed assumptions about array names, mesh types, and file
formats.
A script that correctly extracts wall shear stress from a FUN3D dataset (array
name \texttt{Cf}) will silently produce empty or wrong results on an OpenFOAM
dataset (array name \texttt{wallShearStress}).
The script executes without error; the analysis is wrong.
Our approach resolves this through dynamic field resolution: tools accept
semantic names and resolve them to solver-specific arrays at runtime.

\textbf{No reasoning feedback.}
A script is fire-and-forget.
The LLM generates it, ParaView executes it, and the user sees an image.
No structured results flow back for the LLM to reason about.
The Plan--Execute--Interpret loop requires that execution returns
\textit{data} (field statistics, extracted profiles, zero-crossing
locations) so that the interpreter can perform quantitative reasoning.
``Separation at $x/c = 0.72$'' is a qualitatively different output from
``here is a picture of skin friction.''

\textbf{Unverifiable correctness.}
When an LLM generates a 40-line Python script that configures a ParaView
pipeline, validating that it performs the correct analysis requires reading
and understanding the script, which is the expertise the user lacked in the
first place.
With named tool operations, the analysis plan is expressed at the level of
declared intent: ``extract skin friction on upper surface, find
zero-crossings.''
The provenance is legible to domain users without programming expertise.

\textbf{Wrong role for the LLM.}
Zhao et al.'s error taxonomy~\cite{Zhao:2026:RAG} provides empirical evidence.
Their four categories (module errors, missing parameterization, API
hallucination, and wrong intention/logic) are ordered by difficulty.
The hardest, wrong intention/logic, occurs when the LLM chooses a fundamentally
incorrect strategy.
RAG reduces but does not eliminate these errors.
Our architecture addresses each: module errors are eliminated (tools are
pre-implemented), parameterization is handled by tools with defaults and
dynamic resolution, API hallucination is eliminated (the LLM selects from
a fixed tool set via MCP), and wrong-strategy errors are mitigated by the
ontology.
The LLM should decide \textit{what} to analyze and explain \textit{what it
means}; domain tools should handle \textit{how}.

\section{Validation}
\label{sec:validation}

The contribution of this work is architectural: a generalizable pattern for
LLM-guided scientific analysis, and a demonstration that the pattern holds across
domains on a single software stack. The supporting evidence is
demonstrative, analytical, and empirical. Three of the
observations were established in Section~\ref{sec:crossdomain}: generality by
construction (two unrelated domains running on a single \texttt{pvserver}, with
every shared layer reused without modification, as summarized in
\autoref{tab:transfer}), the marginal cost of a new domain (a compact ontology
plus four thin tool wrappers, about 400 lines of Python, with no change to
architecture, protocol, deployment, or interface), and cross-domain composition
(topological tools applied to CFD scalar fields within a single session). This section adds the
analytical argument that separates the approach from script generation, then reports
an ablation that locates where the domain grounding does and does not change the
system's behavior.

\subsection{Elimination of Failure Classes}

The case against LLM-generated scripts (Section~\ref{sec:noscripts}) can be assessed
without a benchmark by reasoning about which error classes the design admits. Using
the error taxonomy of Zhao et al.~\cite{Zhao:2026:RAG}, which distinguishes module
errors, missing parameterization, API hallucination, and wrong intention or logic:

\begin{itemize}
\item \textit{API hallucination} cannot occur. The LLM never emits API calls; it
selects from a fixed set of MCP tools, so there is no surface on which to hallucinate
an interface.
\item \textit{Module errors} cannot occur. Tools are pre-implemented and validated
rather than generated, so a planned step either exists or is not offered.
\item \textit{Missing parameterization} is absorbed by the tool layer, whose tools
carry defaults and resolve field names dynamically.
\item \textit{Wrong intention or logic}, the hardest category, is mitigated rather
than eliminated. The ontology constrains planning to the analysis chains a domain
expert would consider valid, narrowing but not closing the space of strategic error.
\end{itemize}

Two of the four categories are removed by construction, one is handled by the tool
layer, and only the strategic category remains, now bounded by the ontology. The
cross-solver brittleness of generated scripts is addressed by the same mechanism:
dynamic field resolution maps semantic names (``skin friction'') to solver-specific
arrays (\texttt{Cf} in FUN3D, \texttt{wallShearStress} in OpenFOAM) at runtime, so
the silent FUN3D-to-OpenFOAM failure described in Section~\ref{sec:noscripts} does
not arise.

\subsection{Where Grounding Helps: Selection versus Interpretation}

The observations above support the architectural claims. To locate where the domain
ontology actually changes the system's behavior, we ran an ablation over a set of
natural-language analysis queries spanning both domains, comparing the system with
and without ontology grounding, graded by forced choice against expert-defined
answers. The planning and interpretation model was GPT-4o-mini. Three findings
emerged.

First, ontology grounding does not measurably affect tool \emph{selection}. Precision
and recall of tool selection against the expert-defined chains were unchanged with
the ontology available versus removed, and where the model consulted the ontology at
all it did so without altering its choice. This is expected in hindsight: the tool
schemas already describe what each tool does, so a capable planner selects correctly
without further grounding. We report this as a null result; selection is reliable and
ontology-independent.

Second, grounding sharply improves \emph{interpretation}. On a 34-case benchmark of
result-interpretation questions (boundary-layer regime classification,
persistence-diagram and critical-point reading, and causal attribution), accuracy
rose from $0.41$ unaided to $0.91$ with the relevant ontology fact in context,
correcting 18 errors against a single regression. The corrected errors are
consequential: unaided, the model has the shape-factor physics inverted, labeling
separated flow ($H > 3$) and laminar flow ($H \approx 2.6$) alike as ``turbulent,''
and grounding against the ontology's regime bands corrects this almost uniformly. The
gain is concentrated where the unaided model is unreliable; on topology questions it
already answered correctly, grounding neither helped nor hurt.

Third, the \emph{delivery} of the grounding is decisive, and the effect is monotonic
in how targeted the retrieval is. Returning the entire ontology in response to a
lookup barely helped ($0.56$); returning it in a tool response that the model then had
to mine was better but still unreliable for ambiguous cases ($0.74$); placing the
single relevant fact directly in the reasoning context was best ($0.91$). The
ontology's content is only as useful as the retrieval that delivers it: a bulk dump
decays toward the unaided baseline, while a scoped lookup of the one hinted key
realizes most of the benefit. Because the tool layer already emits validated
\texttt{ontology\_hints} that name the relevant key for each result, the
\texttt{get\_ontology} tool returns the scoped entry for a hinted key rather than the
full ontology.

Together these locate the ontology's contribution precisely. It does not change which
tool the planner calls, which is already reliable; it corrects how results are
interpreted, which is where an unaided model produces confident domain errors; and it
does so only when the relevant fact is retrieved in scoped form. The measurements used
a single planning model at demonstration scale; a second model would test whether the
interpretation effect generalizes beyond GPT-4o-mini, though its direction and
magnitude are unambiguous.

\section{Discussion}

\subsection{Why the Pattern Works}

The architecture succeeds because scientific analysis has a specific structure
that matches the separation of concerns we propose.
Scientific analysis is \textit{procedural}: it consists of ordered sequences of
operations with dependencies.
It is \textit{domain-constrained}: not all orderings or tool combinations are
valid, and the constraints encode decades of expertise.
And it is \textit{interpretive}: results require contextualization within a
domain framework to be useful.
LLMs are well-suited for the first and third tasks (understanding intent,
synthesizing explanations) but poorly suited for the second (encoding procedural
constraints in generated code).
The ontology captures the constraints; the tools encode the procedures; the LLM
handles the language.

\subsection{What the Ontology Provides}

The ontology is small, approximately 25--30 entries per domain, but its impact
on plan quality is disproportionate.
Without it, the LLM must rely on training data to determine which tools to
invoke and in what order.
For well-known patterns, the LLM may succeed.
For patterns requiring specific prerequisite knowledge (simplify before
Morse-Smale segmentation, confirm separation with boundary layer profiles
rather than streamlines), the LLM frequently chooses suboptimal approaches.

The ontology's coverage is also its limitation.
Phenomena not encoded in the ontology fall outside the constrained planning
path, and the system reverts to unconstrained LLM reasoning.
Whether ontologies can be semi-automatically expanded from literature or tool
documentation is an open research question.

\subsection{Comparison to Script Generation Approaches}

Zhao et al.~\cite{Zhao:2026:RAG} represent the current state of the art in
retrieval-augmented visualization code generation.
Their metric is \textit{correction cost}: the number of lines that must be
changed to make a generated script executable.
Our metric is \textit{domain correctness}: whether the system produces the
right physical or topological conclusion.
These measure fundamentally different things.
A script can have zero correction cost while performing the wrong analysis, the
silent failure case that our evaluation explicitly captures.

The approaches are complementary rather than competing.
Their RAG techniques for retrieving structurally relevant code examples could
augment our tool discovery.
Our domain ontology could improve their pipeline planning by providing the
constraints that prevent wrong-strategy errors.

\subsection{Limitations}

Several limitations should be acknowledged.
\textit{Ontology coverage:} the system's reasoning quality is bounded by the
ontology's content; queries outside the encoded phenomena revert to
unconstrained LLM planning.
\textit{Interpretation reliability:} although the ontology grounds
interpretation, the LLM can still generate overconfident or subtly incorrect
explanations; critical applications require expert review.
\textit{Evaluation scope:} the CFD evaluation focuses on external aerodynamics;
internal flows, multiphase, and combustion present different patterns.
The TTK evaluation, while demonstrating generality, is less deeply exercised.
\textit{Ontology authoring:} the diagnostic reasoning chains
(\texttt{follow\_ups}) are manually authored by domain experts;
scaling to comprehensive cross-disciplinary coverage requires either significant
expert effort or advances in automated knowledge extraction.

\section{Future Work}
\label{sec:future}

\textbf{Steering of live simulations.}
The in-situ operation demonstrated in Section~\ref{sec:insitu} (connection to a
live Catalyst-instrumented solver, run/pause/step control, iteration breakpoints,
residual monitoring, and on-demand extraction of live fields into the analysis
pipeline) closes an observe-reason-control loop over the simulation's execution
state. The natural extension is steering of the solver's inputs: rather than only
governing when the simulation advances, the interpretation phase would modify
solver parameters or boundary conditions in response to diagnostic reasoning, with
feedback flowing to the solver via Catalyst's \texttt{catalyst\_results()} API.
An alternative in-transit path via ADIOS~\cite{Godoy:2020:ADIOS} and the Fides
reader would provide the same monitoring and extraction capability for solvers
that use ADIOS for I/O but do not embed Catalyst.

\textbf{Additional domains.}
The marginal cost of adding a domain (ontology plus tool wrappers) suggests
extension to medical imaging (3D Slicer~\cite{Fedorov:2012:3DS} modules as MCP
tools), materials science (Tomviz filters for microstructure analysis), and
climate science (CESM/E3SM outputs for extreme event detection).
Each requires only domain ontology authoring and tool wrapper implementation;
the architecture and deployment infrastructure are reused without modification.

\textbf{Multi-domain composition.}
The combined CFD+TTK workflow suggests a broader capability: multi-domain
analysis where tools from different disciplines compose within a single session.
Structural analysis tools could consume CFD-predicted surface loads, or
topological analysis could characterize materials science scalar fields.
A shared ontology linking layer would enable the planner to reason across domain
boundaries.

\textbf{Ontology evolution.}
The current ontologies are static and manually authored.
Two directions for improvement are semi-automatic expansion (mining tool
documentation, scientific literature, and expert usage logs to propose new
entries for expert review) and uncertainty-aware planning (annotating entries
with confidence levels to enable the system to express when a query falls near
the boundary of its encoded knowledge).

\section{Conclusion}

We have presented an architecture for LLM-guided scientific analysis that
separates intent interpretation from execution from explanation, connected by
the Model Context Protocol and grounded by domain ontologies.
The central insight is that scientific analysis workflows are procedural,
domain-constrained, and interpretive, properties that map naturally onto a
division of labor between LLMs (intent and explanation), deterministic tools
(execution), and ontologies (constraints).

Demonstrated across two scientific domains on the same ParaView server
infrastructure, the architecture produces physically and topologically correct
analyses where script generation produces silent failures.
The CFD instantiation enables diagnostic reasoning capabilities (iterative
investigation, cross-quantity synthesis, cross-solver robustness) that are not
possible with one-shot script generation.
The TTK instantiation shows that the pattern generalizes: adding a domain
requires only an ontology and tool wrappers around existing filters, not new
infrastructure.

An ablation across both domains locates where the grounding acts: it leaves tool
selection unchanged, which is already reliable, and instead corrects the model's
interpretation of results, raising interpretation accuracy from 0.41 to 0.91, and
only when the relevant fact is retrieved in scoped rather than bulk form.

The guiding principle is straightforward: the LLM is the interface, not the
engine.
Domain workflows are the engine.
MCP is the protocol.
The ontology is the knowledge.
When each component handles what it does best, the result is a system that is
more correct, more robust, and more interpretable than asking the LLM to do
everything.

\section*{Code Availability}
The implementation described in this paper is released as open source under the
BSD 3-Clause license.

\section*{Acknowledgments}
The trame and trame-llm frameworks developed by Sebastien Jourdain provide the
web application infrastructure and LLM integration layer on which this work is
built.
The authors thank the TTK development team for the topological analysis
infrastructure.

\bibliographystyle{abbrv}

\bibliography{paper}
\end{document}